\documentclass[12pt]{article}
\usepackage{amsmath,amssymb,amsfonts,color,graphicx,cite,color}
\usepackage{orcidlink}
\usepackage{latexsym}
\usepackage[normalem]{ulem}
\usepackage{epsfig,psfrag,rotating,soul}
\usepackage{rotfloat}
\input paperdef

\evensidemargin \oddsidemargin
\allowdisplaybreaks

\begin{document}
\thispagestyle{empty}

\def\thefootnote{\fnsymbol{footnote}}

\begin{flushright}
\mbox{}
IFT--UAM/CSIC--25-037 
\end{flushright}

\vspace{0.5cm}

\begin{center}

\begin{large}
\textbf{Towards a Refined Understanding of Non-holomorphic Soft}
\\[2ex]
\textbf{SUSY-Breaking Effects on the Higgs Boson Mass Spectra}
\end{large}

\vspace{1cm}

{\sc 
M.~Rehman\orcidlink{0000-0002-1069-0637}$^{1}$%
\footnote{email: m.rehman@comsats.edu.pk}%
~and S.~Heinemeyer\orcidlink{0000-0002-6893-4155}$^{2}$%
\footnote{email: Sven.Heinemeyer@cern.ch}%
}

\vspace*{.7cm}
{\sl
${}^1$Department of Physics, Comsats University Islamabad, 44000
  Islamabad, Pakistan \\[.1em] 
${}^2$Instituto de F\'isica Te\'orica, (UAM/CSIC), Universidad
  Aut\'onoma de Madrid, Cantoblanco, 28049 Madrid, Spain
}
\end{center}

\vspace*{0.1cm}

\begin{abstract}
\noindent

We study the impact of the non-holomorphic (NH) soft supersymmetry-breaking terms \( T_{33}^{\prime D} \) and \( \mu^\prime \), which introduce additional SUSY-breaking effects beyond the holomorphic structure of the superpotential, on the Higgs boson mass spectrum in the NH Minimal Supersymmetric Standard Model (NHSSM).  
The term \( T_{33}^{\prime D} \) modifies the scalar bottom-quark mass matrix and Higgs couplings, while \( \mu^\prime \) affects the mass matrices of charginos and neutralinos.  
In our analysis, we incorporate constraints from charge- and color-breaking (CCB) minima where we find that a portion of the parameter space is excluded by these constraints. Focusing on the allowed parameter space, the NH contributions to the  light  $\cp$-even Higgs boson mass, \( M_h \), from \( \mu^\prime \) and \( T_{33}^{\prime D} \) can reach up to \( 1.4 \gev \) and \( 90 \mev \), respectively.  
For the heavy $\cp$-even Higgs boson mass, \( M_H \), and the charged Higgs boson mass, \( M_{H^{\pm}} \), these contributions can be substantially larger in certain regions of the parameter space, reaching up to \( 44 \gev \) for \( M_H \) and \( 42 \gev \) for \( M_{H^{\pm}} \) due to \( \mu^\prime \), and up to \( 60 \gev \) due to \( T_{33}^{\prime D} \) for both \( M_H \) and \( M_{H^{\pm}} \).  
These corrections are significantly larger than the expected future experimental precision for Higgs boson masses and should therefore be considered in precision analyses for future experiments.

\end{abstract}

\def\thefootnote{\arabic{footnote}}
\setcounter{page}{0}
\setcounter{footnote}{0}

\newpage



\section{Introduction}
\label{sec:intro}

The Minimal Supersymmetric Standard Model (MSSM)~\cite{Fayet:1974pd,Fayet:1976et,Fayet:1977yc, Nilles:1983ge, Haber:1984rc,Barbieri:1987xf} is an important theoretical framework for studying and exploring physics beyond the Standard Model (SM). The introduction of supersymmetry (SUSY) in the SM helps address some fundamental questions in particle physics, such as the hierarchy problem and the nature of dark matter. The MSSM provides natural candidates for dark matter in the form of the lightest supersymmetric particle (LSP), which is typically stable and weakly interacting. The MSSM requires two Higgs doublets, leading to additional Higgs bosons compared to the SM, including two charged Higgs bosons \( H^\pm \), one \(\cp\)-odd Higgs boson \( A \), and two neutral \(\cp\)-even Higgs bosons \( h \) and \( H \). The ratio of the vacuum expectation values of these doublets is defined by the parameter \( \tan\beta := v_2/v_1 \), where typically values ranging from 1 to 50 are assumed. These additional Higgs bosons offer distinctive signatures that can be explored at particle colliders.

Despite extensive searches, the large hadron collider (LHC) has not yet observed any direct evidence of SUSY particles nor for any other beyond the SM (BSM) particle.  The LHC experiments, particularly ATLAS and CMS, have been able to set increasingly stringent lower limits on the masses of SUSY particles. These mass bounds have ruled out certain regions of parameter space and have put pressure on some traditional MSSM scenarios. The hierarchy problem, which pertains to the fine-tuning required in the Higgs boson mass, motivated the introduction of SUSY. However, the absence of sparticles detection and the lack of significant deviations in Higgs boson properties from the SM~\cite{Sekmen:2022vzu} have led researchers to reconsider the naturalness argument for SUSY or to look for interactions beyond to the MSSM to explain the LHC results.

The inclusion of non-holomorphic (NH) terms allows for additional interactions and couplings in the theory~\cite{Girardello:1981wz, Bagger:1993ji}, which can have various implications for the particle spectrum, phenomenology, and experimental observables~\cite{Chattopadhyay:2017qvh, Un:2023wws, Chattopadhyay:2018tqv, Chattopadhyay:2019ycs, Chattopadhyay:2022ecq, Chakrabortty:2011zz, Israr:2024ubp,Rehman:2024tdr, Israr:2025cfd}. This extended model is known as the non-holomorphic supersymmetric standard model (NHSSM). Previous studies~\cite{Jack:1999ud, Jack:1999fa, Jack:2004dv, Cakir:2005hd, Sabanci:2008qp, Un:2014afa}, have examined the effects of these NH terms, particularly in the context of the Higgs boson sector. A more recent work~\cite{Chattopadhyay:2016ivr} investigated the impact of NH soft SUSY-breaking (SSB) terms on the mass of the lightest $\cp$-even Higgs boson $M_h$. They specifically considered the effects of these terms on the scalar top masses, which influence the left-right mixing parameter $X_t$. However, as discussed in \citere{Rehman:2022ydc}, it is important to note that changes in the holomorphic SSB terms, specifically the trilinear Higgs-stop coupling $T_{33}^{U}$, can emulate the observed effects attributed to the NH terms. For a given value of $T_{33}^{\prime U}$, the NH trilinear coupling of scalar top quark, the coupling $T_{33}^{U}$ can be adjusted to yield the same scalar top mass. Consequently, an observed scalar top mass spectrum corresponds to a continuous range of combinations of $T_{33}^{U}$ and $T_{33}^{\prime U}$ while keeping other SSB parameters and the Higgs mixing parameter $\mu$ fixed. Therefore, an analysis that solely varies $T_{33}^{\prime U}$, leading to shifts in the scalar top masses, can not be considered realistic.

Since the NH SSB terms also affect the couplings of the Higgs bosons to the scalar fermions, it is crucial to consider all potential effects simultaneously to accurately determine the genuine impact of the NH terms. This comprehensive approach is necessary to fully understand the implications of NH effects within the NHSSM. In our previous analysis~\cite{Rehman:2022ydc}, we investigated the impact of the parameter $T_{33}^{\prime U}$ on the masses of the Higgs bosons while ensuring that no effects on the scalar top masses and mixings arise. The influence of $T_{33}^{\prime U}$ on the mass of the lightest $\cp$-even Higgs boson $M_h$ was found to be relatively small, in contrast to the claim in \citere{Chattopadhyay:2016ivr}, with deviations only reaching a few $\mev$. However, the effects of $T_{33}^{\prime U}$ on the heavy $\cp$-even  Higgs boson mass $M_H$ and the charged Higgs boson mass $M_{H^{\pm}}$ were significant, leading to deviations of several $\gev$. In the mass matrix of the scalar top sector, the coupling $T_{33}^{\prime U}$ is divided by the parameter $\tb$, resulting in relatively minor effects~\cite{Rehman:2022ydc}. However, in the scalar bottom sector, the NH trilinear coupling $T_{33}^{\prime D}$ is multiplied by $\tb$, possibly enhancing its effect on the Higgs boson masses, particularly $M_h$. Additionally, the NH parameter $\mu^\prime$, which appears in the neutralino and chargino mass matrices, can have important implications for the Higgs boson phenomenology, particularly when considering constraints related to dark matter and low fine-tuning issue\cite{Un:2023wws}.

In this study, we analyze the impact of NH SSB terms, namely $T_{33}^{\prime D}$ and $\mu^\prime$, on the mass spectra of the Higgs bosons complementing our analysis in \citere{Rehman:2022ydc}, where we focused on $T_{33}^{\prime U}$ contributions. We calculate the corresponding one-loop corrections to the NHSSM Higgs-boson masses in the Feynman diagrammatic approach. The analysis of the effects on the Higgs-boson masses was performed with the help of {\tt SPheno}~\cite{Porod:2003um} source code specifically designed for the NHSSM. This source code was obtained through the Mathematica package {\tt SARAH}~\cite{Staub:2009bi,Staub:2010jh,Staub:2012pb,Staub:2013tta,Staub:2015kfa}, which provides a framework for generating the necessary model files and calculations for particle physics phenomenology. In the numerical analysis, similar to our approach in \citere{Rehman:2022ydc}, we ensured that the physical masses of the particles entering the loop contributions are either fixed, or vary only slightly, below possible future experimental precisions. This is crucial for a realistic set-up.

The structure of the paper is outlined as follows: initially, we describe the key characteristics of the NHSSM in \refse{sec:model_NHSSM}. The computational framework is discussed in \refse{sec:CompSetup} and our  numerical findings are presented in \refse{sec:NResults}. Lastly, our conclusions are summarized in \refse{sec:conclusions}.

\section{Model set-up}
\label{sec:model_NHSSM}

The superpotential within the MSSM is required to be holomorphic, leading to the parametrization of the SSB sector primarily with holomorphic operators. Nevertheless, it is possible to extend the MSSM by incorporating R-Parity violating and/or NH terms in the SSB sector~\cite{Girardello:1981wz, Bagger:1993ji, Chakrabortty:2011zz}. In its simplest form, the SSB sector of the MSSM can include the following terms: 
\begin{eqnarray}
\label{NonH-TrilinearTerms}
-\cL_{\rm soft}^{\rm NH}&=&T_{ij}^{^\prime D}h_2 {\tilde d}_{Ri}^*{\tilde q}_{Lj}
+T_{ij}^{^\prime U}h_1 {\tilde u}_{Ri}^*{\tilde q}_{Lj}
+T_{ij}^{^\prime E}h_2 {\tilde e}_{Ri}^*{\tilde l}_{Lj}
+\mu^{\prime} {\tilde h}_1 {\tilde h}_2
\end{eqnarray}
Here, ${\tilde q}_{L_i}$ (${\tilde l}_{L_i}$) are the left-handed squark (slepton) doublet fields while  ${\tilde u}_{R_i}$, ${\tilde d}_{R_i}$ and ${\tilde e}_{R_i}$ are the right-handed up-type squark, down-type squark and charged slepton singlet fields respectively with $i,j=1,2,3$ representing the generation index. The $h_1$, $h_2$ represent two Higgs doublets and \( \tilde{h}_1 \) and \( \tilde{h}_2 \) are the fermionic components (Higgsinos) of the down-type and up-type Higgs superfields, respectively. The $\mu^{\prime}$ represents the NH higgsino mass term. The $T_{ij}^{^\prime U}= Y_{ij}^{U} A_{ij}^{^\prime U}$, $T_{ij}^{^\prime D}= Y_{ij}^{D} A_{ij}^{^\prime D}$, and $T_{ij}^{^\prime E}= Y_{ij}^{E} A_{ij}^{^\prime E}$ denote the NH trilinear coupling matrices for up-type squarks, down-type squarks, and charged sleptons, respectively with $Y_{ij}$ representing the corresponding Yukawa coupling. It's important to note that these terms are not necessarily related to the holomorphic trilinear soft terms. One approach is to consider a scenario where these non-holomorphic trilinear terms are assumed to be equal to the holomorphic trilinear couplings as a "boundary condition" at the Grand Unified Theory (GUT) scale, especially in models like the Constrained MSSM. However, it's essential to acknowledge that due to renormalization group equation running effects, the NH trilinear terms will evolve differently, resulting in completely different values from the holomorphic ones, as discussed in \citere{Un:2014afa}. Therefore, it is reasonable to treat the NH trilinear terms as independent quantities, although they are expected to be of similar magnitudes to the usual trilinear couplings. This approach is meaningful when comparing predictions of the NHSSM with experimental results.

In the presence of the NH trilinear terms, the down-type squarks mass matrix can be written as  
\begin{equation}
M_{\tilde{D}}^{2}=\left(
\begin{array}
[c]{cc}%
m_{\tilde{D}_{LL}}^{2} & m_{\tilde{D}_{LR}}^{2}\\[.5em]
m_{\tilde{D}_{LR}}^{2\dag} & m_{\tilde{D}_{RR}}^{2}%
\end{array}
\right) \label{fermion mass matrix}%
\end{equation}
with%
\begin{align}
m_{\tilde{D}_{LL}}^{2}  & =m_{\tilde{Q}}^{2}+M_{Z}^{2}\cos2\beta\left(  I_{3}%
^{f}-Q_{f}s_{W}^{2}\right)  +m_{d}^{2}\nonumber\\
m_{\tilde{D}_{RR}}^{2}  & =m_{\tilde{D}}^{2}+M_{Z}^{2}\cos2\beta Q_{f}s_{W}^{2}+m_{d}^{2}\nonumber\\
m_{\tilde{D}_{LR_{ii}}}^{2}   & = \frac{v_1}{\sqrt{2}} \left(T_{ii}^{D}-(\mu Y_{ii}^{D}+T_{ii}^{^\prime D}) \tb\right).
\label{mass terms}%
\end{align}
Here, $I_{3}^{f}$ represents the weak isospin of fermions, $Q_{f}$ stands for the electromagnetic charge, and $m_{d}$ denotes the mass of the SM down-type quarks whereas $m_{\tilde{Q}}$, and $m_{\tilde{D}}$ correspond to the masses of the left-handed squark doublet and right-handed down-type squark singlet. $\MZ$ and $\MW$ correspond to the masses of the $Z$ and $W$ bosons, while $s_W$ is defined as the square root of $1 - c_W^2$, where $c_W = \MW/\MZ$ and $T_{ii}^{D}$ represent the holomorphic trilinear couplings corresponding to down-type squarks, while \( \mu \) denotes the holomorphic Higgsino mass parameter. It should be noted that because of the different combination of fields, the
NH trilinear couplings $T_{ii}^{^\prime D}$ receive the additional factors of $\tb$. 
The NH higgsino mass parameter $\mu^{\prime}$ mentioned in \refeq{NonH-TrilinearTerms} modifies the neutralino and chargino mass
matrices given by   
\BE
\renewcommand{\arraystretch}{1.2}
{\bf Y} = \MLv M_1 & 0 & -M_Z\sw\Cb & M_Z\sw\Sbe \\ 0 & 
          M_2 & M_Z\cw\Cb & -M_Z\cw\Sbe \\
          -M_Z\sw\Cb & M_Z\cw\Cb & 0 & -(\mu+\mu^\prime) \\ M_Z\sw\Sbe & 
          -M_Z\cw\Sbe & -(\mu+\mu^\prime) & 0 \MR~,
\label{Eq:Neutralino-Mass-Matrix}
\end{equation}

\BE
{\bf X} = \ML M_2 & \sqrt2\, M_W\, \Sbe \\[1ex] \sqrt2\, M_W\, 
          \Cb & (\mu+\mu^\prime) \MR~.
\label{Eq:Chragino-Mass-Matrix}          
\end{equation}
 
As discussed before, the the NH trilinear terms also modify the
Higgs-sfermion-sfermion couplings. Here we show the couplings of the
lightest Higgs boson $h$ to the down-type squarks.
\begin{align}
C(h,\tilde{d}_{i}^{s},\tilde{d}_{j}^{t}) &  =\frac{-ie\delta_{ij}}{6M_{W}%
c_{W}s_{W}c_{\beta}}\Big[3c_{W}m_{d_{i}}\{s_{\alpha}\frac{T_{ii}^{D}}{Y_{ii}^{D}}+(\mu
+\frac{T_{ii}^{\prime D}}{Y_{ii}^{D}})c_{\alpha}\}U_{s,1}^{\tilde{d},i}U_{t,2}^{\tilde{d}%
,i}\nonumber\\
&  +\{6s_{\alpha}c_{W}m_{d_{i}}^{2}-M_{W}M_{Z}s_{\alpha+\beta}c_{\beta
}(3-2s_{W}^{2})\}U_{s,1}^{\tilde{d},i}U_{t,1}^{\tilde{d},i}\nonumber\\
&  +\{6s_{\alpha}c_{W}m_{d_{i}}^{2}-2M_{W}M_{Z}s_{\alpha+\beta}c_{\beta}%
s_{W}^{2}\}U_{s,2}^{\tilde{d},i}U_{t,2}^{\tilde{d},i}\nonumber\\
&  +3c_{W}m_{d_{i}}\{s_{\alpha}\frac{T_{ii}^{D}}{Y_{ii}^{D}}+(\mu+\frac{T_{ii}^{\prime D}}{Y_{ii}^{D}})c_{\alpha
}\}U_{s,2}^{\tilde{d},i}U_{t,1}^{\tilde{d},i}\Big]\label{ChSqSq}%
\end{align}
The coupling of the charged Higgs boson $H^{-}$ to up-type and
down-type squarks is given by 
\begin{align}
C(H^{-},\tilde{u}_{i}^{s},\tilde{d}_{j}^{t}) &  =\frac{ieV_{ij}^{\rm CKM}}%
{2M_{W}s_{W}s_{\beta}}\Big[m_{u_{i}}U_{s,2}^{\tilde{u},i}U_{t,1}^{\tilde{d}%
,j}\{\frac{T_{ii}^{U}}{Y_{ii}^{U}}+(\mu+\frac{T_{ii}^{\prime U}}{Y_{ii}^{U}})t_{\beta}\}\nonumber\\
&  +m_{u_{i}}m_{d_{j}}U_{s,2}^{\tilde{u},i}U_{t,2}^{\tilde{d},j}(1+t_{\beta
}^{2})+U_{s,1}^{\tilde{u},i}U_{t,2}^{\tilde{d},j}m_{d_{j}} t_{\beta}\{\frac{T_{ii}^{D}}{Y_{ii}^{D}}t_{\beta}+(\mu+ \frac{T_{ii}^{\prime D}}{Y_{ii}^{D}})\}\nonumber\\
&  +U_{s,1}^{\tilde{u},i}U_{t,1}^{\tilde{d},j}\{m_{u_{i}}^{2}-t_{\beta}%
(M_{W}^{2}s_{2\beta}-m_{d_{j}}^{2})t_{\beta}\}\Big]
\label{CHpSqSq}
\end{align}
Here $i,j$ are the generation indices (we assume flavor
conservation throughout the paper), $U_{s,s'}^{\tilde{u},i}$
($U_{t,t'}^{\tilde{d},j}$) is the 
$2\times 2$ rotation matrix for up-type (down-type)
squarks that diagonalizes the scalar top (scalar bottom) mass matrix, and we use the shorthand notation $s_x, c_x, t_x$ for
$\sin x$, $\cos x$, $\tan x$, respectively,
where $\alpha$ is the $\cp$-even Higgs mixing angle.
The couplings of the $\cp$-even heavy Higgs boson $H$
to the down-type squarks can be obtained by replacing $c_{\alpha}
\rightarrow s_{\alpha}$, $s_{\alpha} \rightarrow -c_{\alpha}$ and
$s_{\alpha+\beta} \rightarrow -c_{\alpha+\beta}$ in \refeq{ChSqSq}. 
It is interesting to observe that the $T_{ii}^{\prime D}$ enter differently
into the scalar bottom masses and into the trilinear Higgs-scalar bottom
couplings. In the mass matrix, it appears with a factor of $\tb$, whereas the holomorphic trilinear term does not include this factor. Similarly, in the couplings, $T_{ii}^{\prime D}$ is multiplied by $c_{\alpha}$, while the trilinear term is multiplied by $s_{\alpha}$. This will be crucial for our numerical analysis, see the
discussion in \refse{sec:NResults}.


\section{Higher order corrections in the NHSSM Higgs sector}
\label{sec:CompSetup}

\subsection{Tree-level structure and higher-order corrections}

The MSSM (and thus the NHSSM) Higgs-boson sector consist of two Higgs
doublets and predicts the existence of five physical
Higgs bosons, the light and heavy $\cp$-even $h$ and $H$, the $\cp$-odd $A$,
and a pair of charged Higgs bosons, $H^\pm$, where we assume $\cp$ conservation throughout the paper. At the tree-level the Higgs
sector is described with the help of two parameters: the mass of the
$A$~boson, $\MA$, and $\tb = v_2/v_1$, the ratio of the two vacuum
expectation values. 
The tree-level relations and in particular the tree-level masses receive
large higher-order corrections, see, 
e.g., \citeres{Draper:2016pys,Slavich:2020zjv} and references
therein.

The lightest MSSM Higgs boson, with mass $\Mh$, can be interpreted as
the new state discovered at the LHC around
$\sim 125 \gev$~\cite{Heinemeyer:2011aa}. 
The present experimental uncertainty at the LHC for $\Mh$, 
is about~\cite{ParticleDataGroup:2024cfk},
\begin{align}
\de\Mh^{\rm exp,today} \sim 110 \mev~.
\end{align}
This can possibly be reduced to the level of 
\begin{align}
\de\Mh^{\rm exp,future} \lsim 20 \mev
\end{align}
at HL-LHC~\cite{Cepeda:2019klc} and future $e^+e^-$ colliders (where we take ILC\cite{Bambade:2019fyw,deBlas:2019rxi} numbers as concrete example).
Similarly, for the masses of the heavy neutral Higgs 
$\MH$, an uncertainty at the $1\%$ level 
could be expected at the LHC~\cite{Gennai:2007ys}.

\medskip
At the tree-level, the masses of the $\cp$-even Higgs bosons are given by

\BE
m_{(H,h),\text{tree}}^{2}=\frac{1}{2}\left[  M_{A}^{2}+M_{Z}^{2}\pm
\sqrt{(M_{A}^{2}+M_{Z}^{2})^{2}-4M_{A}^{2}M_{Z}^{2}\cos2\beta}\right]
\end{equation}

However, the tree-level masses, receive large higher-order corrections~\cite{Draper:2016pys,Slavich:2020zjv}. In the Feynman diagrammatic (FD) approach employed in our current calculation\footnote{Precise calculations of $\Mh$ go beyond the FD approach described here \cite{Slavich:2020zjv}. However, the NHSSM corrections that are the focus of this article are obtained at the one-loop level. Since we are interested only in their relative effects, the restriction to the one-loop corrections in the FD approach is sufficient.}, the higher-order corrections are included in the matrix $\Delta_{\rm Higgs}$ given by
\BE
\left(\Delta_{\rm Higgs}\right)^{-1}
= - i \ML p^2 -  \mHtree^2 + \hSi_{HH}(p^2) &  \hSi_{hH}(p^2) \\
     \hSi_{hH}(p^2) & p^2 -  \mhtree^2 + \hSi_{hh}(p^2) \MR~.
\label{higgsmassmatrixnondiag}
\end{equation}

The masses of the $\cp$-even Higgs bosons are determined by finding the poles of the propagator matrix for $(h,H)$. Determining the poles of the matrix $\Delta_{\rm Higgs}$ in \refeq{higgsmassmatrixnondiag} is equivalent to solving
the equation
\begin{equation}
\left[p^2 - \mhtree^2 + \hSi_{hh}(p^2) \right]
\left[p^2 - \mHtree^2 + \hSi_{HH}(p^2) \right] -
\left[\hSi_{hH}(p^2)\right]^2 = 0\,.
\label{eq:proppole}
\end{equation}

Likewise, concerning the charged Higgs sector, the charged Higgs mass is deduced from the position of the pole in the charged Higgs propagator \cite{Frank:2013hba} which can be obtained by solving the equation
\noindent \begin{equation}
p^{2}-m^{2}_{H^{\pm},{\rm tree}} +
\hat{\Sigma}_{H^{-}H^{+}}\left(p^{2}\right)=0.
\label{eq:proppolech}
\end{equation}

\begin{figure}[htb!]
\centering
\includegraphics[width=12cm,height=8cm,keepaspectratio]{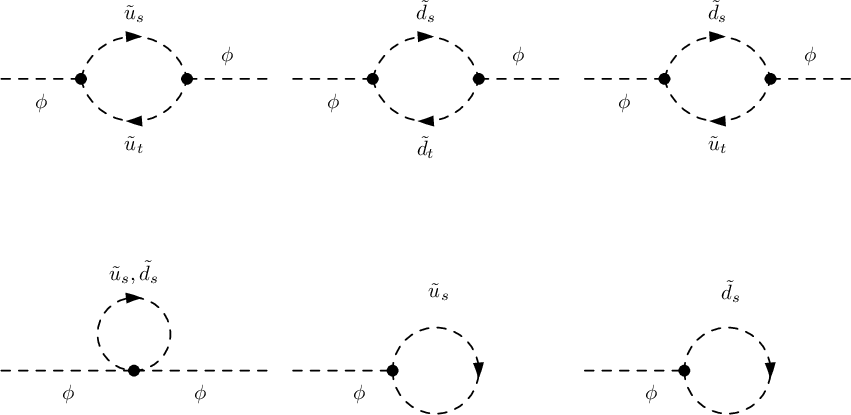}
\caption{
Generic Feynman diagrams for the Higgs boson self-energies and
tadpoles. $\phi$ denotes any of the Higgs bosons, $h$, $H$, $A$ or 
$H^\pm$; $u$ stand for $u,c,t$; $d$ stand for $d,s,b$; $\tilde u_{s,t}$
and $\tilde d_{s,t}$ are the six mass 
eigenstates of up-type and down-type squarks, respectively.} 
\label{FDHSelf}
\end{figure}

The (renormalized) self-energies of the Higgs bosons as described in \refeqs{eq:proppole} and \ref{eq:proppolech} can be computed at the $n$-loop level through explicit FD calculations of the relevant loop diagrams. As previously mentioned, our focus in this study will be primarily on the one-loop corrections originating from the bottom/sbottom sector and $\mu^\prime$. Generic Feynman diagrams that involve NH couplings are shown in the \reffi{FDHSelf}. Here, we restrict our analysis to quark/squark contributions only, as these are the only sectors affected by $T_{33}^{\prime D}$ or $\mu^{\prime}$. The parameter $\mu^{\prime}$ does not enter the interaction vertices but contributes solely to the chargino and neutralino mass matrices. Consequently, by fixing the combination $\mu + \mu^{\prime}$ to a constant value---an approach adopted in this analysis---the neutralino and chargino sectors remain unaffected by variations in $\mu^{\prime}$.

The NH SSB parameters in the NHSSM play a crucial role in the one-loop computation of the various (renormalized) self-energies and tadpoles of the Higgs bosons. As previously discussed, these parameters impact scalar fermion masses as well as the Higgs-sfermion-sfermion coupling, as depicted in \refeqs{ChSqSq}-(\ref{CHpSqSq}), which will be particularly significant in our analysis. 


\subsection{Constraints}
\label{sec:constraints}

In general, the contributions of $T_{33}^{\prime D}$ to the Higgs masses are anticipated to be large due to the multiplication of the $T_{33}^{\prime D}$ coupling by $\tb$. In our specific setup, where we maintain the same physical mass spectrum, the value of $T_{33}^{ D}$ could potentially become quite large due to the fixed nature of left-right mixing parameter of the scalar bottom quark $X_{b}$ (see below for an exact definition). This large $T_{33}^{ D}$ value may result in the violations of constraints related to charge and color breaking minima (CCB). The CCB constraints in the framework of NHSSM were analyzed in \citeres{Beuria:2017gtf, Chattopadhyay:2019ycs} and are given as 
\begin{align}
\left(  \sum_{i=d,s,b}(T_{ii}-\mu y_{ii}-T_{ii}^{^{\prime}})\right)  ^{2}  &
<\left(  \frac{g_{1}^{2}+g_{2}^{2}}{2}+5\sum_{i}y_{ii}^{2}\right) \nonumber \\
& \times \left(  \frac{1}{2}\sum_{i}(M_{\tilde{Q}_{ii}}^{2}+M_{\tilde{D}_{ii}}%
^{2})+m_{H_{u}}^{2}+m_{H_{d}}^{2}+2\left\vert \mu\right\vert ^{2}-2B_{\mu
}\right).
\label{Eq-CCB-Constraints}
\end{align}
Here, \( M^2_{Q_{ii}} \) and \( M^2_{D_{ii}} \) denote the soft mass-squared terms for the left-handed squark doublets and right-handed down-type squarks of the \( i \)th generation, respectively. The bilinear soft term in the Higgs potential, \( B_{\mu} \), is given by
\begin{equation}
B_{\mu} = \frac{M_A^2 \sin 2\beta}{2},
\end{equation}
while the up- and down-type Higgs boson mass parameters, \( m^2_{H_u} \) and \( m^2_{H_d} \), are calculated as  
\begin{align}
m^2_{H_u} &= \frac{M_A^2 \cos^2\beta + \frac{1}{2} M_Z^2 \cos 2\beta - \mu^2}{1 + \tan^2\beta}, \\
m^2_{H_d} &= \frac{M_A^2 \sin^2\beta - \frac{1}{2} M_Z^2 \cos 2\beta - \mu^2}{1 + \cot^2\beta},
\end{align}
where \( M_A \) is the pseudoscalar Higgs mass, and \( g_1 \) and \( g_2 \) are the gauge couplings of the \( U(1) \) and \( SU(2) \) gauge groups, respectively. 
For our numerical analysis, we wrote a private code based on \refeq{Eq-CCB-Constraints} to calculate the CCB constraints. 

\subsection{Computational workflow}
\label{sec:strategy0}

To calculate the NH contributions to the Higgs boson masses, we start by generating the NHSSM source code for \texttt{SPheno}~\cite{Porod:2003um} using the \texttt{SARAH} package~\cite{Staub:2009bi, Staub:2010jh, Staub:2012pb, Staub:2013tta, Staub:2015kfa}. \texttt{SARAH} is a Mathematica-based tool that automates the generation of source code for various new physics models, including the NHSSM. By using the NHSSM model file defined in \texttt{SARAH}, we produce the required code, which is then used to calculate the SUSY spectrum including Higgs boson masses, and other relevant observables. For the numerical analysis, we utilize an additional tool, \texttt{SARAH Scan and Plot (SSP)}~\cite{Staub:2011dp}, where we define the input parameters. \texttt{SSP} generates an SLHA file to be used by \texttt{SPheno}, which is then called to calculate the particle spectrum and other observables. The output from \texttt{SPheno} is read by \texttt{SSP}, which subsequently uses this data for plotting and further analysis.

\subsubsection{$T_{33}^{\prime D}$ analysis}
To extract pure $T_{33}^{\prime D}$ term contributions, it is important that the physical masses and mixing of the scalar bottom remain the same. This can be achieved by fixing the left-right mixing parameter $X_b$. Consequently, for our numerical evaluation we fix $X_b$ and calculate the value of $T_{33}^{D}$ for the given value of $T_{33}^{\prime D}$ using the expression  
\begin{equation}
\Xb = T_{33}^{D} - (\mu Y_{33}^{D} + T_{33}^{\prime D})\tb,    
\end{equation}
where $T_{33}^{\prime D}$ is taken as a free parameter in a specific range i.e. $-100 \gev<T_{33}^{\prime D}<100 \gev$. 
Concerning the numerical evaluation, for a given value of
$T_{33}^{D}$ in the MSSM and $T_{33}^{\prime D}$ in the 
NHSSM a new value of $T_{33}^{D}$ is calculated such that
\begin{equation}
\Xb^{\rm MSSM} = T_{33}^{D ~{\rm MSSM}}- \mu Y_{ij}^{D}\tb    
\end{equation}
and
\begin{equation}
\Xb^{\rm NHSSM} = T_{33}^{D ~{\rm NHSSM}}- (\mu Y_{ij}^{D}+T_{33}^{\prime D}) \tb    
\end{equation}
are identical (yielding the same values for the sbottom masses and mixings). Using $T_{33}^{D ~{\rm MSSM}}$ and
$T_{33}^{\prime D}$ the NH contribution to the Higgs-boson masses are calculated numerically. 

\subsubsection{\boldmath{$\mu^\prime$} analysis}

For the $\mu^\prime$ contributions, we set all NH trilinear couplings to zero (including $T_{33}^{\prime D}$)  and fix $\mu + \mu^\prime$ to the value of $\mu$ specified in the corresponding scenario. We then vary $\mu^{\prime}$ within the range of $-2000 \gev$ to $+2000 \gev$. 
It should be noted that the parameter $\mu^{\prime}$ does not directly appear in the couplings (see \refeq{ChSqSq} and \refeq{CHpSqSq}); rather, it enters only through the neutralino and chargino mass matrices (see \refeq{Eq:Neutralino-Mass-Matrix} and \refeq{Eq:Chragino-Mass-Matrix}). The observed effects on the Higgs boson masses arise because we impose the condition $\mu + \mu^{\prime} = \text{constant}$. Varying $\mu^{\prime}$ leads to a corresponding change in $\mu$, which affects not only the couplings of squarks and sleptons to the Higgs bosons but also the sfermion mass matrices. Hence, the effects originate from the variation in $\mu$.  

Contrary to the approach adopted for $T_{33}^{\prime D}$ analysis, two approaches are followed for the $\mu^\prime$ analysis . In the first approach, the holomorphic trilinear terms $T_{ii}^{f}$ are kept constant, so any change in $\mu$ leads to a corresponding change in $X_f$. As a result, both the Higgs-sfermion couplings and the sfermion masses and mixing are affected, leading to large contributions.  

In contrast, in the second approach, $X_f$ is kept fixed, and the change in $\mu$ is compensated by an appropriate adjustment of the holomorphic trilinear terms $T_{ii}^{f}$. In this case, the contributions arise solely from the variation in $\mu$ within the Higgs-sfermion couplings, resulting in smaller contributions.

In the benchmark scenarios used for our analysis (defined in the following section), the holomorphic trilinear couplings are taken to be equal, i.e., $A_t = A_b = A_{\tau}$. For each point P1–P4 (defined in the next section), these terms are determined from the corresponding value of $X_t$ in that scenario. Using the relation
\begin{equation}
    X_t = A_t - \frac{\mu}{\tan\beta},
\end{equation}
we compute the holomorphic trilinear couplings $A_t = A_b = A_{\tau}$, along with the corresponding soft terms \( T_{33}^{U} \), \( T_{33}^{D} \), and \( T_{33}^{E} \). In the first approach, these values are fixed, while \( X_t \), \( X_b \) and \( X_\tau \) vary with changing \( \mu^{\prime} \). In the second approach, \( X_t \), \( X_b \), and \( X_\tau \) are initially determined for \( \mu^{\prime} = 0 \) based on the corresponding values of \( T_{33}^{U} \), \( T_{33}^{D} \), and \( T_{33}^{E} \) at that point. Subsequently, \( X_t \), \( X_b \), and \( X_\tau \) are kept constant by adjusting \( T_{33}^{U} \), \( T_{33}^{D} \), and \( T_{33}^{E} \) as \( \mu' \) varies.

\section{Numerical Results}
\label{sec:NResults}
 
\subsection{Input parameters}
\label{sec:input-para}

For our numerical analysis, we opted for three scenarios initially introduced in \citere{Bahl:2018zmf}. These scenarios, denoted as $M_{h}^{125}$, $M_{h}^{125}(\tilde{\tau})$, and $M_{h}^{125}(\tilde{\chi})$, provide insight into different aspects of Higgs boson phenomenology within the MSSM. They are characterized by specific SSB parameters and exhibit good agreement with experimental data from the LHC. Each scenario features a $\cp$-even scalar with a mass approximately around 125 GeV, resembling the SM Higgs boson. In these scenarios, the SSB mass $M_{\tilde f}$ for the first two generations is taken as $2 \tev$, while the holomorphic trilinear SSB terms for these generations are set to zero. The remaining input parameters are provided in ~\refta{tab:input-parameters}~\cite{Bahl:2018zmf}.

\begin{table}[h] \centering
\begin{tabular}{lccr}
\hline\hline
& $M_{h}^{125}$  & $M_{h}^{125}(\tilde{\tau})$ & $M_{h}^{125}%
(\tilde{\chi})$ \\\hline\hline
$m_{\tilde{Q}_{3}},m_{\tilde{U}_{3}},m_{\tilde{D}_{3}}$ & $1500$ & $1500$ &
$1500$\\
$m_{\tilde{L}_{3}},m_{\tilde{E}_{3}}$ & $2000$ & $350$ & $2000$\\
$\mu$ & $1000$ & $1000$ & $180$\\
$M_{1}$ & $1000$ & $180$ & $160$\\
$M_{2}$ & $1000$ & $300$ & $180$\\
$M_{3}$ & $2500$ & $2500$ & $2500$\\
$X_{t}$ & $2800$ & $2800$ & $2500$\\
$A_{\tau}$ & $A_t$ & $800$ & $A_t$\\
\hline
$m_{\tilde b_{1}},m_{\tilde b_{2}}$ & 1339,1662   & 1339,1662
 & 1358,1646  \\  
\hline\hline
\end{tabular}
\caption{Selected scenarios in the MSSM parameter space, taken from
  \citere{Bahl:2018zmf}. All the dimensionful quantities are in $\gev$.}%
\label{tab:input-parameters}
\end{table}%
In Table~\ref{tab:input-parameters}, $m_{\tilde{Q}_{3}}$, $m_{\tilde{U}_{3}}$, and $m_{\tilde{D}_{3}}$ correspond to the masses of the third generation squark doublet, up-type squark singlet, and down-type squark singlet, respectively. Furthermore, $m_{\tilde{L}_{3}}$ and $m_{\tilde{E}_{3}}$ denote the masses of the third generation left-handed slepton doublet and right-handed slepton singlet, respectively. The parameters $M_{1}$, $M_{2}$, and $M_{3}$ denote the gaugino masses, while $\mu$ represents the standard Higgs mixing parameter. Here, $A_\tau$ represent the holomorphic trilinear coupling such that $A_\tau=T_{33}^{E}/ Y_{33}^{E}$. In these scenarios, the holomorphic trilinear couplings of the third generation are set equal, i.e., \( A_t = A_b = A_\tau \), where \( A_t = T_{33}^{U}/Y_{33}^{U} \) and \( A_b = T_{33}^{D}/Y_{33}^{D} \), except in the \( M_{h}^{125}(\tilde{\tau}) \) scenario, where \( A_\tau = 800 \,\text{GeV} \). For each scenario, we investigate four different combinations of $M_A$ and $\tb$, taking into account the latest experimental limits for MSSM
Higgs-boson searches~\cite{ATLAS:2020zms,CMS:2022goy}:
\begin{align*}
\rm P1 &:~\MA = 1500 \gev,~ \tb = 7 \\
\rm P2 &:~\MA = 2000 \gev,~ \tb = 15 \\
\rm P3 &:~\MA = 2500 \gev,~ \tb = 30 \\
\rm P4 &:~\MA = 2500 \gev,~ \tb = 45 
\end{align*}
and define
\begin{equation}
\begin{aligned}
    \delta M_h & \equiv M_h^{\text{NHSSM}} - M_h^{\text{MSSM}}, \\
    \delta M_H & \equiv M_H^{\text{NHSSM}} - M_H^{\text{MSSM}}, \\
    \delta M_{H^\pm} & \equiv M_{H^\pm}^{\text{NHSSM}} - M_{H^\pm}^{\text{MSSM}}.
\end{aligned}
\end{equation}
Here $M_{h, H, H^\pm}^{\text{MSSM}}$ represents the values of $M_{h, H, H^\pm}$ when NH terms $T_{33}^{\prime D}$ and $\mu^{\prime}$ are set to zero.

\subsection{NH contributions to the Higgs-boson masses}

\subsubsection{\boldmath{$\mu^\prime$} contributions}

In this section, we present our results for $\mu^\prime$ contributions to Higgs boson masses. As already explained, we fix $\mu+\mu^\prime$ to the value of $\mu$ in the respective scenario. 
This approach will keep the chargino and neutralino phenomenology (masses and mixing) unchanged which will have important consequences for the dark matter relic abundance. For example, we have verified that \( \Omega_{\rm CDM} h^2 \) remains unchanged for all points in the \( M_h^{125} \) and \( M_h^{125}(\tilde{\chi}) \) scenarios. However, it varies in the \( M_h^{125}(\tilde{\tau}) \) scenario. In this case, the charged slepton, i.e., \( \tilde{e}_1 \), becomes the next-to-lightest supersymmetric particle (NLSP), and the change arises due to variations in the masses \( m_{\tilde{e}_{1,2}} \). These mass variations result from changes in \( \mu \), as there is no compensating factor \( \mu' \) in the scalar lepton mass matrix. On the other hand, in the original definition of the benchmark planes in \citere{Bahl:2018zmf}, DM was {\em not} considered as a relevant constraint. The reason is that the parameters affecting the DM relic density have only a small impact on the Higgs-boson sector. Furthermore, a small amount of R-parity violation would fully invalidate any bounds on the DM relic density, while having a negligible impact on the Higgs-sector phenomenology.

Furthermore, it is important to note that point P1 is ruled out by CCB constraints, as the small value of \( M_A \) at this point causes the left-hand side of \refeq{Eq-CCB-Constraints} to dominate. Nevertheless, we show the contributions from P1 for completeness. Also here it should be noted that these constraints were not considered in the original definition of the benchmark planes in \citere{Bahl:2018zmf}.

\begin{figure}[ht!]
\begin{center}
\psfig{file=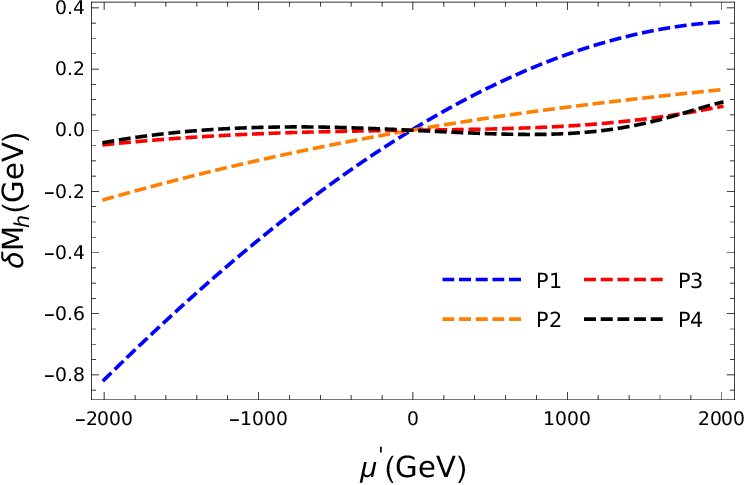  ,scale=0.64,angle=0,clip=}
\psfig{file=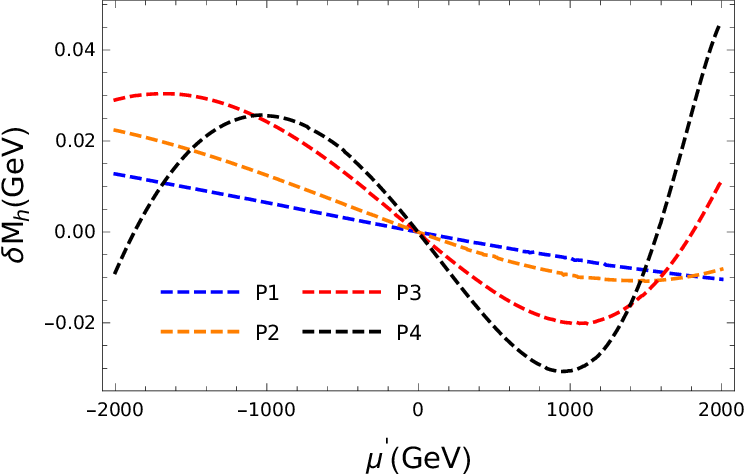  ,scale=0.64,angle=0,clip=}\\
\vspace{0.6in}
\psfig{file=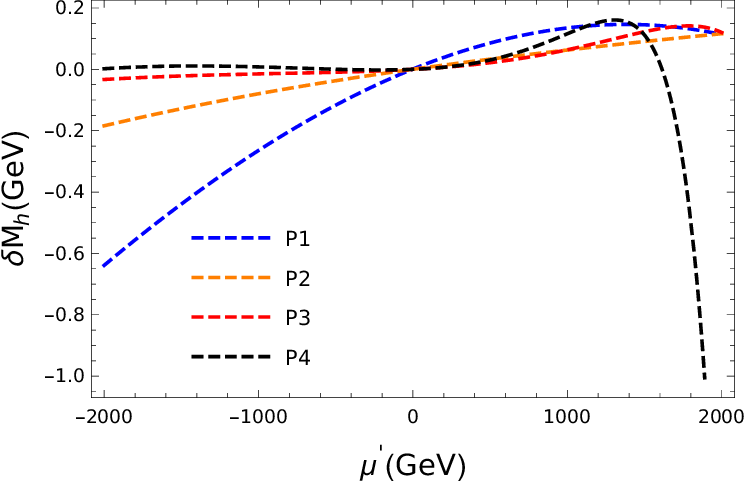  ,scale=0.64,angle=0,clip=}
\psfig{file=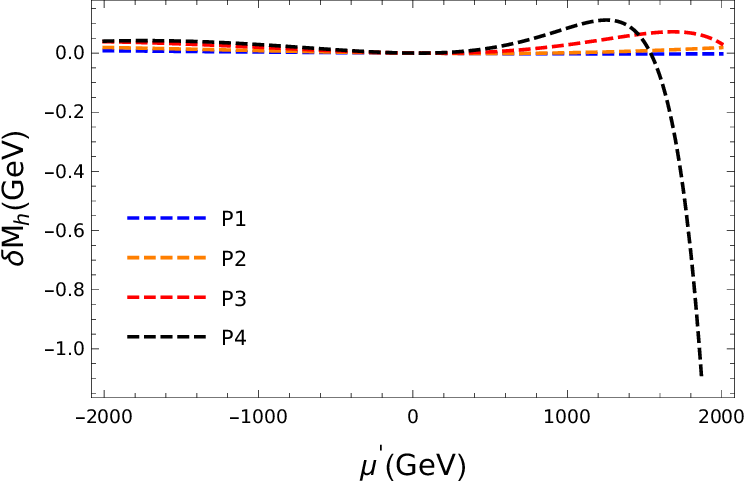  ,scale=0.64,angle=0,clip=}\\
\vspace{0.6in}
\psfig{file=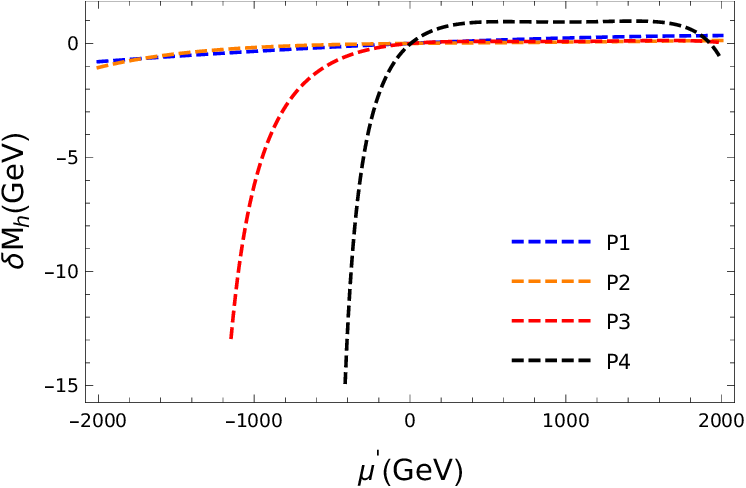  ,scale=0.64,angle=0,clip=}
\psfig{file=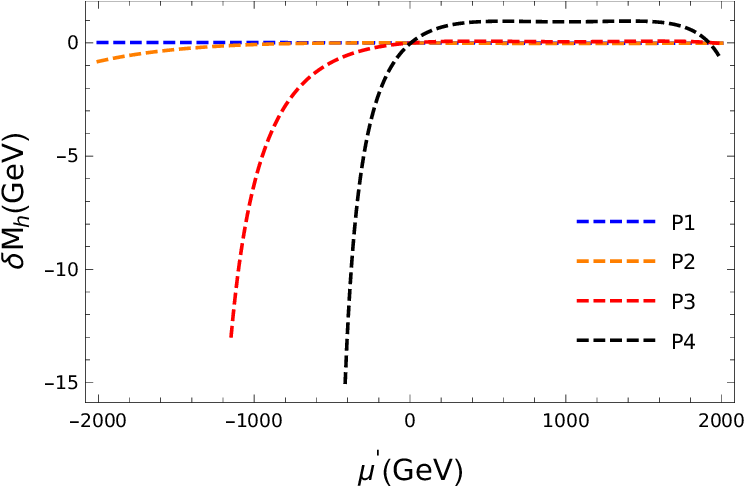  ,scale=0.64,angle=0,clip=}\\
\vspace{0.3in}
\end{center}
\caption{ $\delta M_h$  as a function of $\mu^{\prime}$ for $M_h^{125}$
  (upper row),  $M_h^{125}(\tilde{\chi})$ (middle row) and $M_h^{125}(\tilde{\tau})$ (lower row) scenarios for fixed $T_{ii}^{f}$ (left panel) and fixed $X_f$ (right panel).}
\label{fig:Mup-Mh0}
\end{figure}

In \reffi{fig:Mup-Mh0}, we present our results for the contributions of $\mu^{\prime}$ to $M_h$ in the $M_h^{125}$ (upper row), $M_h^{125}(\tilde{\chi})$ (middle row), and $M_h^{125}(\tilde{\tau})$ (lower row) scenarios. In the left panel, we show the result where $T_{ii}^{f}$ are fixed whereas right panel show the results where $X_f$ is fixed. For cases where the holomorphic trilinear terms $T_{ii}^{f}$ are fixed, the largest contributions occur at lower values of $M_A$ and $\tb$ in $M_h^{125}$. Specifically, at point P1, the contributions can reach up to $+350 \mev$ and $-850 \mev$, while other points exhibit comparatively smaller contributions. In the $M_h^{125}(\tilde{\chi})$ scenario, point P1 contributions can reach up to $+150 \mev$ and $-650 \mev$, whereas contributions from other points remain below $200 \mev$ except for the point P4 which can reach up to $-1 \gev$ for large values of $\mu^{\prime}$.  In contrast, the $M_h^{125}(\tilde{\tau})$ scenario exhibits a different trend, with the largest contributions arising from point P4, reaching up to $-15 \gev$, while point P3 contributes up to $-13 \gev$. For point P4 (P3), values of $\mu^{\prime} < -400 ~(-1200) \gev$ are not allowed, as they lead to negative slepton squared masses. The apparently large contributions arise from the fixed value of the holomorphic trilinear term $A_{\tau}$ in this scenario. Moreover, due to the small values of $m_{\tilde{L}_{3}},m_{\tilde{E}_{3}}$, the  masses of the scalar taus are significantly affected by changes in $\mu$, leading to large contributions. Thus, the large effects should be viewed more as an artifact of the definition of the benchmark scenario.
Previously, we emphasized that pure NHSSM contributions can only be isolated if the masses and mixing angles are held constant. However, in the current approach—where we fix $T_{ii}^{f}$ to a constant value while allowing $X_f$ to vary—the physical mass spectrum of the sfermions is affected. Although the spectrum changes, we have explicitly verified that the variations remain within 2\% in most cases, which is well below the expected experimental uncertainty. An exception arises in the $M_h^{125}(\tilde{\tau})$ scenario, where the variation in slepton masses can be significant. This behavior, however, should be attributed to the specific structure of the scenario, as discussed above.

For cases where $X_f$ is fixed, the contributions remain small. In the $M_h^{125}$ scenario, point P4 exhibits the largest contributions, reaching up to $\pm 45 \mev$. Similarly, in the $M_h^{125}(\tilde{\chi})$ scenario, P4 again shows the largest contributions, reaching up to $-1.4 \gev$, while other points showing only minor contributions.  

In the $M_h^{125}(\tilde{\tau})$ scenario, although the left-right mixing parameters for scalar top and scalar bottom quarks are fixed, the trilinear holomorphic term $T_{33}^{E}$ is also fixed due to the scenario definition. Consequently, the left-right mixing parameter for scalar tau lepton varies, leading to contributions similar to those observed when $X_f$ was not fixed.

\begin{figure}[ht!]
\begin{center}
\psfig{file=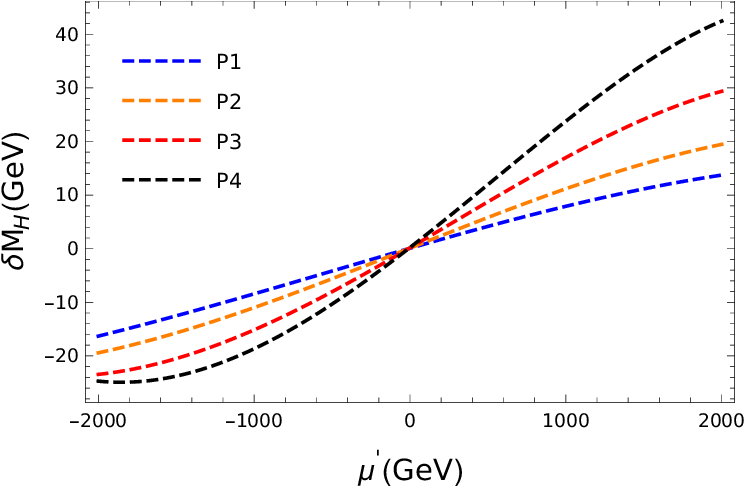  ,scale=0.64,angle=0,clip=}
\psfig{file=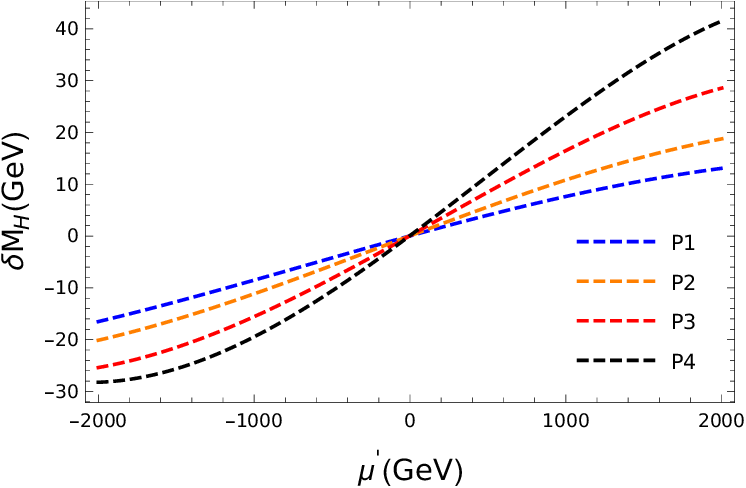  ,scale=0.64,angle=0,clip=}\\
\vspace{0.6in}
\psfig{file=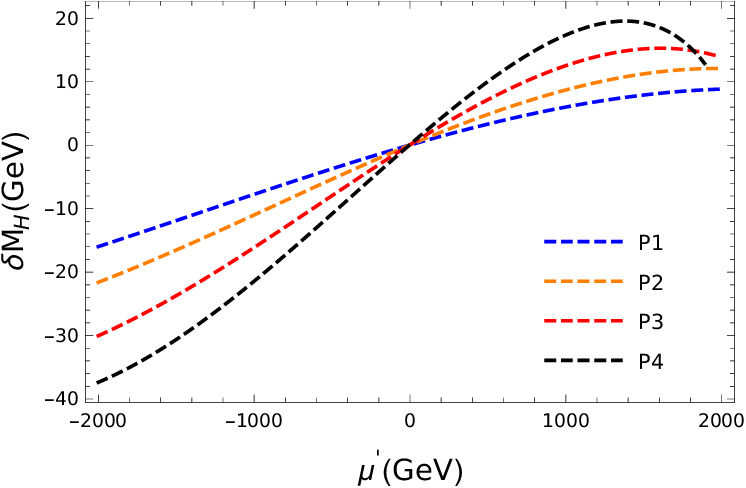  ,scale=0.64,angle=0,clip=}
\psfig{file=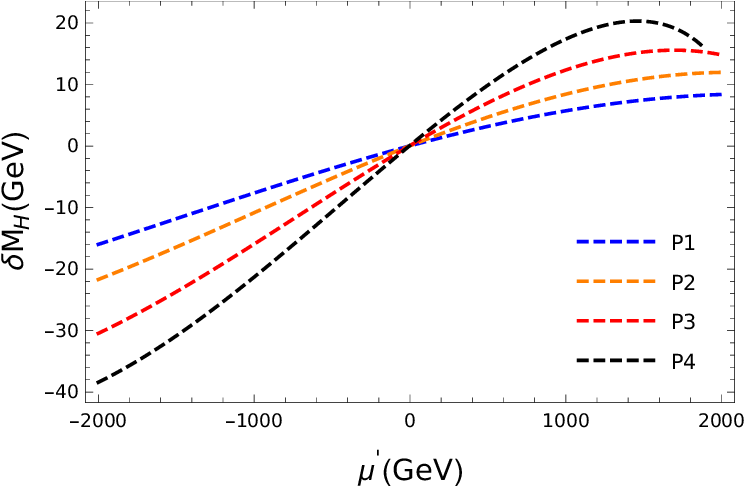  ,scale=0.64,angle=0,clip=}\\
\vspace{0.6in}
\psfig{file=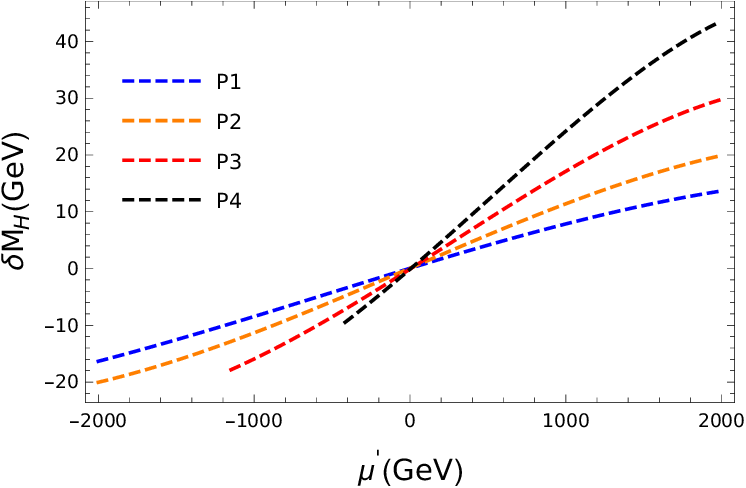  ,scale=0.64,angle=0,clip=}
\psfig{file=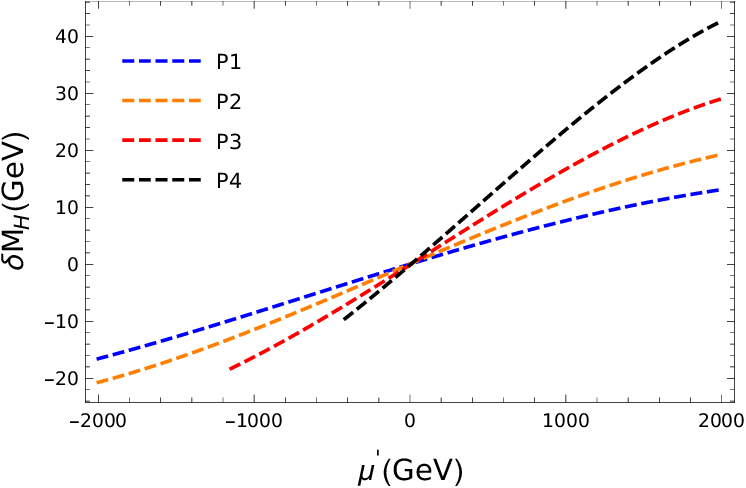  ,scale=0.64,angle=0,clip=}\\
\vspace{0.3in}
\end{center}
\caption{ $\delta M_H$  as a function of $\mu^{\prime}$ for $M_h^{125}$
  (upper row),  $M_h^{125}(\tilde{\chi})$ (middle row) and $M_h^{125}(\tilde{\tau})$ (lower row) scenarios for fixed $T_{ii}^{f}$ (left panel) and fixed $X_f$ (right panel).}
\label{fig:Mup-MHH}
\end{figure}

In \reffi{fig:Mup-MHH}, we present our results for $\delta M_H$ as a function of $\mu^{\prime}$. The plot arrangement follows the same structure as in the previous figure. For the case where the holomorphic trilinear terms $T_{ii}^{f}$ are fixed (left panel), the largest contributions occur at point P4, reaching up to $+44 \gev$ and $-26 \gev$ in the $M_h^{125}$ scenario, and up to $+20 \gev$ and $-40 \gev$ in the $M_h^{125}(\tilde{\chi})$ scenario. In $M_h^{125}(\tilde{\tau})$ scenario, the largest contributions are again for P4 where they reach up to $+44 \gev$. However, for negative values of $\mu^{\prime}$, the contributions from P2 can reach up to $-20 \gev$.  For the case where $X_f$ is fixed (right panel), the overall trend remains the same, though the contributions in some cases are slightly different compared to those in the fixed $T_{ii}^{f}$ scenario.  

\begin{figure}[ht!]
\begin{center}
\psfig{file=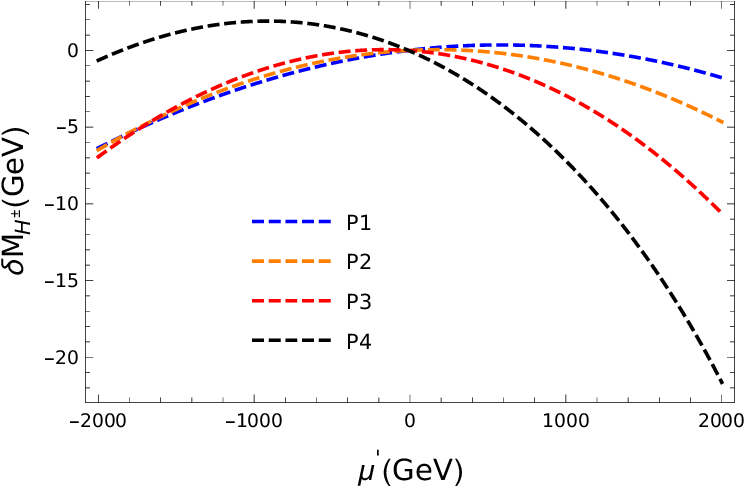  ,scale=0.64,angle=0,clip=}
\psfig{file=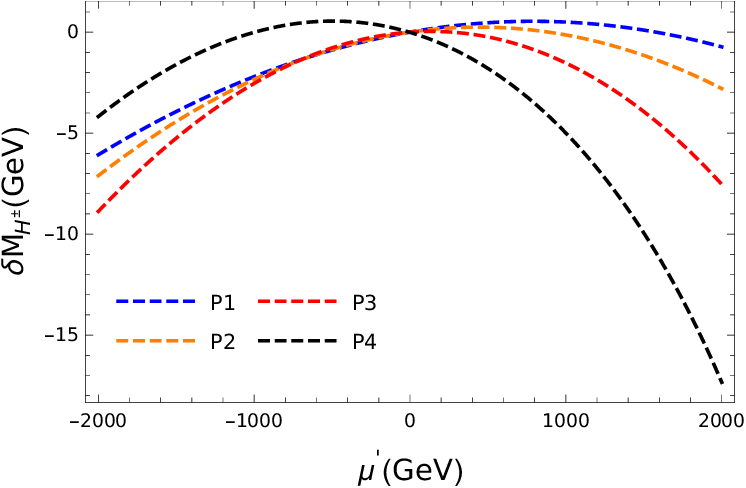  ,scale=0.64,angle=0,clip=}\\
\vspace{0.6in}
\psfig{file=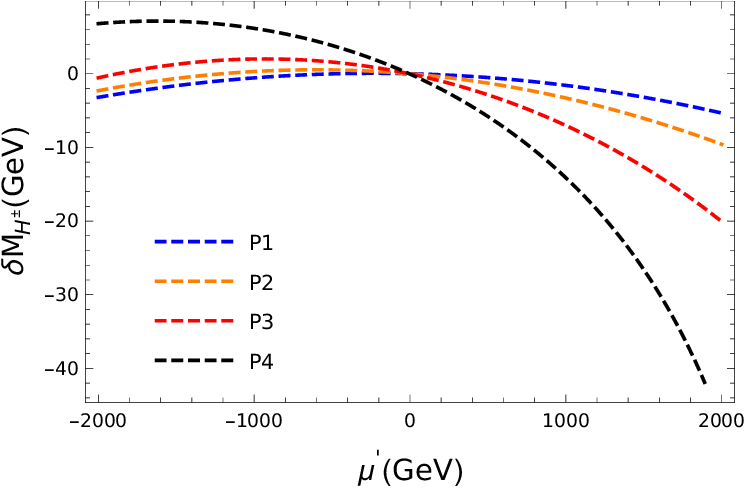  ,scale=0.64,angle=0,clip=}
\psfig{file=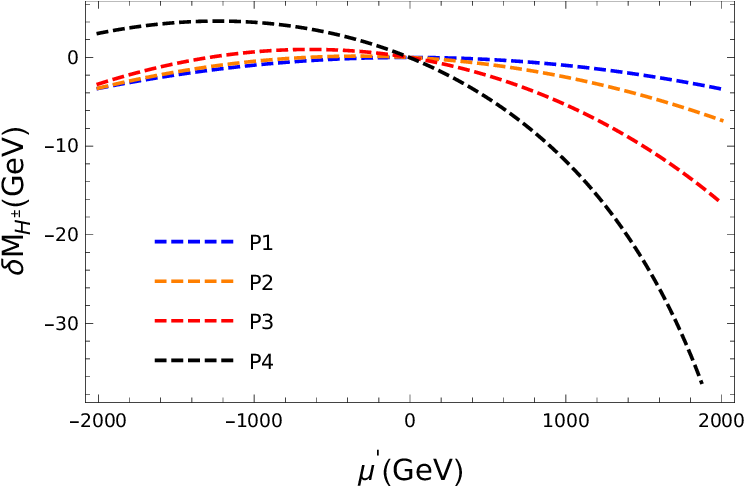  ,scale=0.64,angle=0,clip=}\\
\vspace{0.6in}
\psfig{file=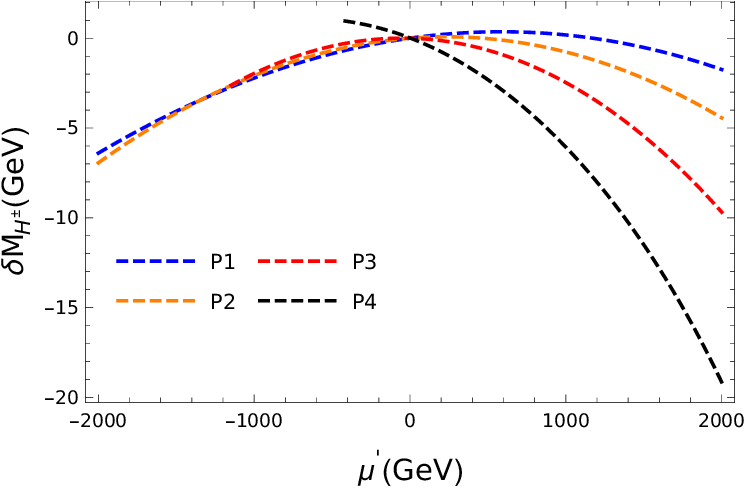  ,scale=0.64,angle=0,clip=}
\psfig{file=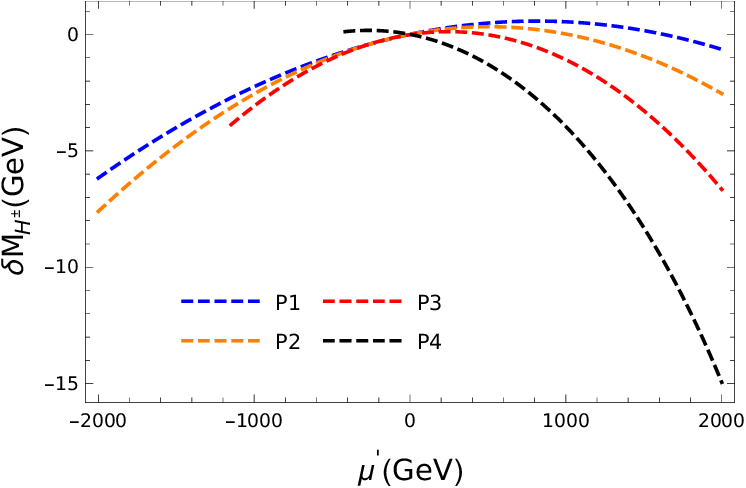  ,scale=0.64,angle=0,clip=}\\
\vspace{0.3in}
\end{center}
\caption{ $\delta M_{H^\pm}$  as a function of $\mu^{\prime}$ for $M_h^{125}$
  (upper row),  $M_h^{125}(\tilde{\chi})$ (middle row) and $M_h^{125}(\tilde{\tau})$ (lower row) scenarios for fixed $T_{ii}^{f}$ (left panel) and fixed $X_f$ (right panel).}
\label{fig:Mup-MHp}
\end{figure}

In \reffi{fig:Mup-MHp}, we show our results for $\delta \MHp$ as a function of $\mu^{\prime}$, with the plot arrangement consistent with the previous figures. In the $M_h^{125}$ scenario with fixed $T_{ii}^{f}$ (left panel), point P4 exhibits the most significant negative contribution, reaching up to $-22 \gev$, while small positive contributions of up to $+2 \gev$ are also observed for this point. Contributions from other points remain within $\pm 10 \gev$ or lower. Similarly, in the $M_h^{125}(\tilde{\chi})$ scenario, P4 again shows the largest contributions, reaching up to $+7 \gev$ and $-44 \gev$. The $M_h^{125}(\tilde{\tau})$ scenario exhibits the most pronounced effects, with point P4 contributing up to $-18 \gev$. For the fixed $X_f$ scenario, the contributions are generally $2$ to $3 \gev$ smaller than those observed in the fixed $T_{ii}^{f}$ case.

\subsubsection{\boldmath{$T_{33}^{\prime D}$} contributions}

In this section, we present our findings regarding the impact of $T_{33}^{\prime D}$ on the Higgs-boson masses $\Mh$, $\MH$, and $\MHp$.  As previously discussed, in these scenarios, the holomorphic trilinear couplings are equal by definition, i.e., \( A_t = A_b \). These couplings, along with the corresponding \( T_{33}^{U} \) and \( T_{33}^{D} \), can be determined from the value of \( X_t \). In \refta{tab:TD}, we show the values of the holomorphic trilinear coupling \( T_{33}^{D} \) for the benchmark points P1 to P4 in the \( M_h^{125} \) and \( M_h^{125}(\tilde{\chi}) \) scenarios. These values correspond to three choices of \( T_{33}^{\prime D} \): $-100 \gev$ (Min), $0$ (Central), and $100 \gev$ (Max). Since the input parameters \( \mu \) and \( X_t \) are identical in the \( M_h^{125} \) and \( M_h^{125}(\tilde{\tau}) \) scenarios, the results for the latter are the same as those in the \( M_h^{125} \) case and are not shown separately.

\begin{table}[h] 
\centering
\setlength{\tabcolsep}{10pt}
\renewcommand{\arraystretch}{1.5}
\begin{tabular}{|l|ccc|ccc|}
\hline\hline
& \multicolumn{3}{c|}{$M_{h}^{125}$} & \multicolumn{3}{c|}{$M_{h}^{125}(\tilde{\chi})$} \\
\hline
Point & Min & Central & Max & Min & Central & Max \\
\hline
P1 & -650 & 50 & 750 & -747 & -48 & 652   \\
P2 & -1451 & 48 & 1549 & -1660 & -160 & 1340  \\
P3 & -2951 & 48 & 3048 & -3370 & -370 & 2630  \\
P4 & -4452 & 48 & 4547 & -5080 & -579 & 3920  \\
\hline\hline
\end{tabular}
\caption{Values of \( T_{33}^{D} \) (in $\gev$) for points P1–P4 in the \( M_h^{125} \) and \( M_h^{125}(\tilde{\chi}) \) scenarios, corresponding to \( T_{33}^{\prime D} = -100 \gev\) (Min), \( T_{33}^{\prime D} = 0 \) (Central), and \( T_{33}^{\prime D} = 100 \gev\) (Max). The values for the \( M_h^{125}(\tilde{\tau}) \) scenario are identical to those of the \( M_h^{125} \) scenario, as the input parameters \( \mu \) and \( X_t \) are the same in both cases.}
\label{tab:TD}
\end{table}

\begin{figure}[ht!]
\begin{center}
\psfig{file=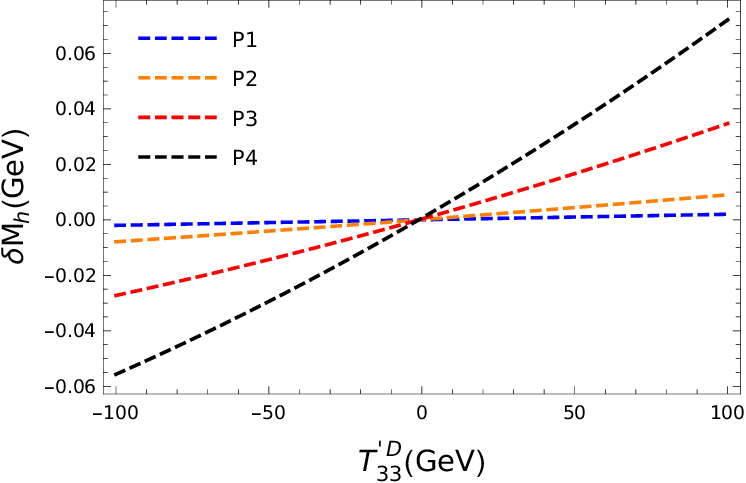  ,scale=0.60,angle=0,clip=}
\hspace{0.5cm}
\psfig{file=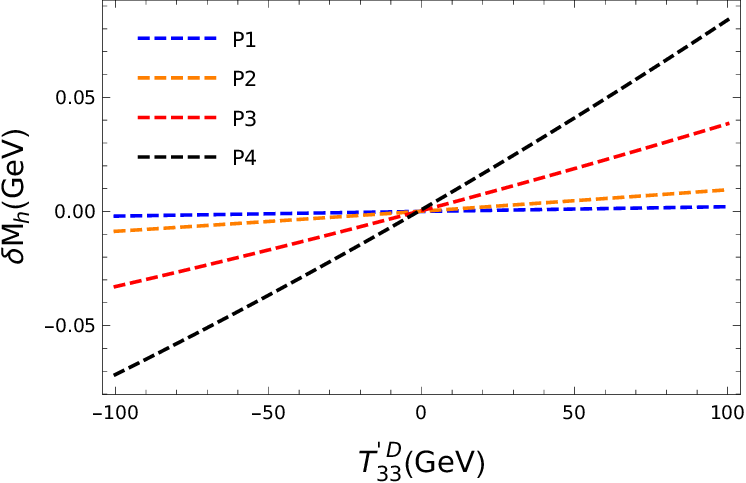  ,scale=0.60,angle=0,clip=}\\
\end{center}
\caption{ $\delta M_h$  as a function of $T_{33}^{\prime D}$ for $M_h^{125}$
  (left plot) and  $M_h^{125}(\tilde{\chi})$ (right plot).}
\label{fig:Mh0-Abp}
\end{figure}

For our numerical evaluation, we vary $T_{33}^{\prime D}$ in the range of $-100 \gev$ to $100 \gev$. In principle, the $T_{33}^{\prime D}$ contributions can be substantial due to the multiplication of the $T_{33}^{\prime D}$ term by $\tb$. 
The plots in \reffi{fig:Mh0-Abp} illustrates the value of $\delta M_h$ as a function of $T_{33}^{\prime D}$. The left plot corresponds to the $M_h^{125}$ scenario, while the right plot shows the results in  the $M_h^{125}(\tilde{\chi})$ scenario. Since the input parameters for the scalar quark sector are the same in $M_h^{125}$ and $M_h^{125}(\tilde{\tau})$, the results for $M_h^{125}(\tilde{\tau})$ align with those of $M_h^{125}$ and are therefore not displayed separately. As anticipated, the most notable effects occur in the P4, characterized by the highest $\tb$ value. For this point, the modifications can be as large as $75 \mev$ for the $M_h^{125}$ scenario and up to $90 \mev$ for the $M_h^{125}(\tilde{\chi})$ scenario.

\begin{figure}[ht!]
\begin{center}
\vspace{0.4in}
\psfig{file=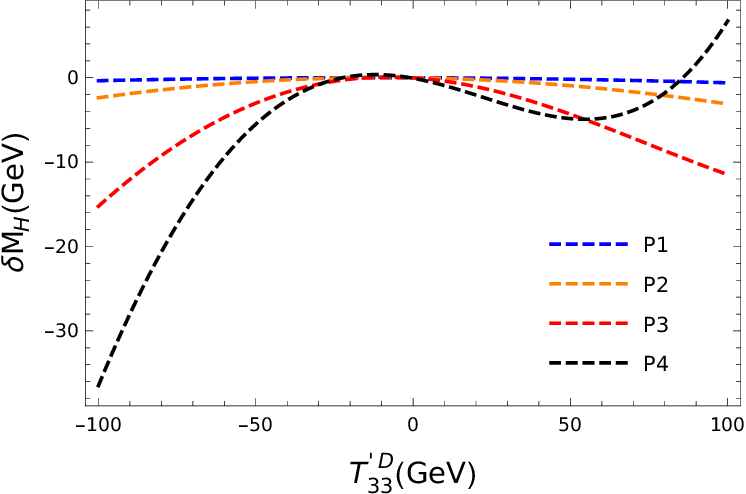  ,scale=0.60,angle=0,clip=}
\hspace{0.5cm}
\psfig{file=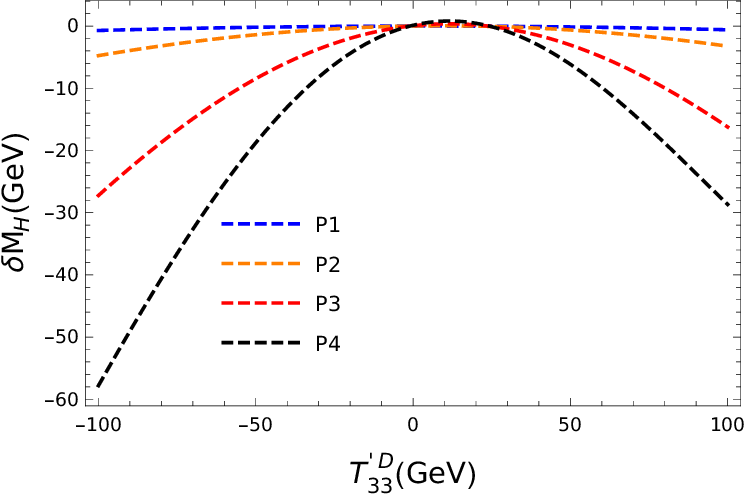  ,scale=0.60,angle=0,clip=}\\
\end{center}
\caption{ $\delta M_H$  as a function of $T_{33}^{\prime D}$ for $M_h^{125}$
  (left plot) and  $M_h^{125}(\tilde{\chi})$ (right plot).}
\label{fig:MHH-Abp}
\end{figure} 

\begin{figure}[ht!]
\begin{center}
\vspace{0.4in}
\psfig{file=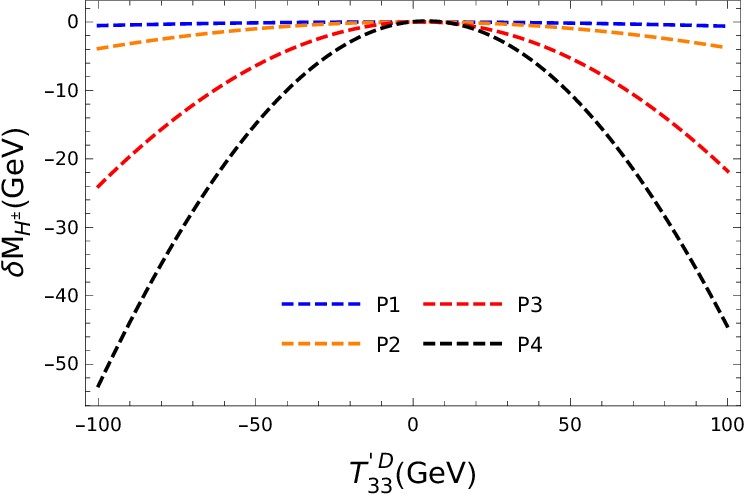  ,scale=0.60,angle=0,clip=}
\hspace{0.5cm}
\psfig{file=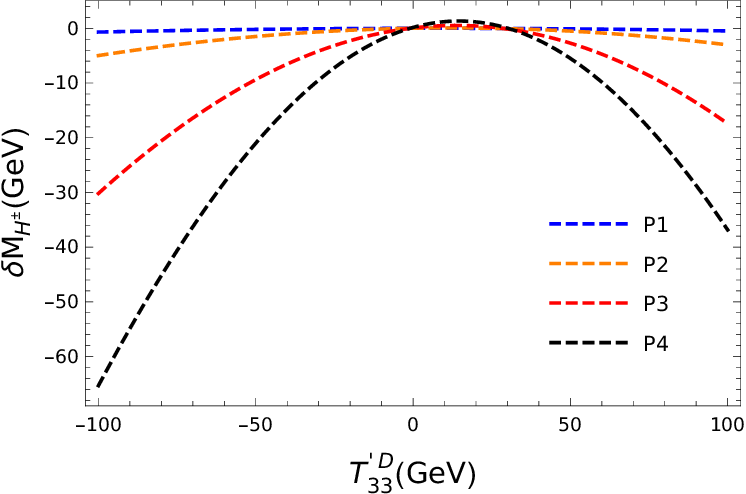  ,scale=0.60,angle=0,clip=}\\
\end{center}
\caption{ $\delta M_{H^\pm}$  as a function of $T_{33}^{\prime D}$ for for $M_h^{125}$
  (left plot) and  $M_h^{125}(\tilde{\chi})$ (right plot).}
\label{fig:MHp-Abp}
\end{figure} 

In Figure \ref{fig:MHH-Abp}, we depict $\delta M_H$ as a function of $T_{33}^{\prime D}$. The layout of plots is the same as in \reffi{fig:Mh0-Abp}. In the context of the $M_h^{125}$ scenario, the most substantial contributions are observed for the point P3 and P4, yielding negative contributions of up to $16 \gev$, however, P4 can result in  negative contribution of up to $38 \gev$. On the other hand, in the $M_h^{125}(\tilde{\chi})$ scenario, the most significant impact emerges from the point P4, resulting in a negative contribution of up to $58 \gev$.

Finally, our findings for $\delta \MHp$ as a function of $T_{33}^{\prime D}$ are presented in Figure \ref{fig:MHp-Abp}. The arrangement of plots follows the same pattern as the previous figures. Once again, the most substantial contributions are associated with the point P4. In the $M_h^{125}$ and $M_h^{125}(\tilde{\chi})$ scenarios, these contributions can extend up to $54 \gev$ and $66 \gev$ respectively.

It should be noted that if the effects arise through mass matrices, then for $T_{33}^{\prime D}$, one would expect significant effects on $M_h$ due to the multiplication of the $T_{33}^{\prime D}$ term by $\tan\beta$ in the mass matrix. However, in our approach, where the mass spectrum is fixed by keeping $X_b$ constant, the situation is reversed. In this case, the multiplication of $T_{33}^{\prime D}$ by $\tan\beta$ causes $T_{33}^{D}$ to become very large. Since this term is multiplied by $c_\alpha$ in the couplings of the $\cp$-even heavy Higgs $H$, the dominant contributions appear in $M_H$.

\section{Conclusions}
\label{sec:conclusions}

In this paper, we have investigated the effect of non-holomorphic (NH) soft
SUSY-breaking terms to the Higgs-boson mass predictions in the NHSSM.
In a previous paper~\cite{Rehman:2022ydc} we concentrated on the effect of the
NH trilinear Higgs-scalar top coupling, $T_{33}^{\prime U}$.
In this paper we complement that work with an analysis of the
NH trilinear Higgs-scalar bottom coupling, $T_{33}^{\prime D}$ and the
NH higgsino mass parameter, $\mu^\prime$.
In order to perform the calculations, using the Mathematica package
{\tt SARAH},  we generated a {\tt Spheno} version including the NH
contributions to the Higgs-boson self-energies. 
This {\tt Spheno} version was then employed for the numerical analysis.

We took particular care to analyze the pure NH contribution. The
NH SSB terms enter into the respective mass matrices, in this case the
scalar bottom and the chargino/neutralino mass matrices. The $T_{33}^{\prime D}$ term modifies both the scalar bottom mass matrix and the Higgs–scalar bottom couplings. In contrast, $\mu^\prime$ appears only in the chargino and neutralino mass matrices, but not in the couplings between the Higgs bosons and the charginos/neutralinos.
An analysis simply varying the NH SSB terms 
thus leads to a shift in the scalar bottom and/or chargino/neutralino 
masses, which should be considered as a different physics scenario, 
as it is expected that the masses and mixing angles of the contributing
SUSY particles are measured in the future (if SUSY is realized).
Consequently, an observed effect from a naive variation of the NH SSB
terms can be 
mimicked by a change in the holomorphic SSB terms. Concretely, 
for each choice of $T_{33}^{\prime D}$ the parameter $T_{33}^D$ 
can be adjusted to yield the same scalar bottom masses. An observed
sbottom mass spectrum thus corresponds 
to a continuous set of combinations of $T_{33}^D$ and $T_{33}^{\prime D}$
(keeping the other SSB and $\mu$ fixed).
An analysis that simply varies $T_{33}^{\prime D}$, resulting in shifts
in the sbottom masses and mixings, 
can thus not be regarded realistic. Therefore, in our analysis,
we required $X_b$ to be constant under a change of $T_{33}^{\prime D}$ by an
adjustment of $T_{33}^D$. In this way, the effect of the NH contributions is
shifted into the Higgs-sbottom couplings and can readily be analyzed.
A similar strategy has been followed for $\mu^\prime$, where we required
that $\mu^\prime + \mu$, as they appear in the chargino/neutralino mass
matrices, remains constant. However, here another subtlety
arises, as $\mu$, but not $\mu^\prime$, also enters into the scalar
fermion mass matrices. Here we used two different options.
In the first approach, the holomorphic trilinear terms are fixed according to the values defined in the respective scenarios. Consequently, variations in $\mu^\prime$, and thereby in $\mu$, lead to changes in the left-right mixing parameters $X_f$. Although the masses and mixing angles in this case change, we have explicitly verified that the resulting shifts remain below 2\% in the majority of cases, staying well within the anticipated experimental uncertainty.
In the second approach, however, the change in $\mu^\prime$ is compensated by a corresponding adjustment in the holomorphic trilinear terms, keeping $X_f$ constant.

For the numerical analysis, we chose three LHC benchmark scenarios
($M_h^{125}$, $M_h^{125}(\tilde\tau)$ and $M_h^{125}(\tilde\chi)$)~\cite{Bahl:2018zmf}.
In each scenario, four combinations of $(\MA, \tb)$ were chosen
that are allowed by current MSSM Higgs-boson searches at the LHC,
$(1000 \gev, 7)$, $(1500 \gev, 15)$, $(2500 \gev, 30)$, $(2500 \gev, 45)$,
called P1, P2, P3, P4, respectively. 
$T_{33}^{\prime D}$ has been varied from $-100 \gev$ to $+100 \gev$,
whereas $\mu^\prime$ was varied from $-2000 \gev$ to $+2000 \gev$.
The relatively small range for $T_{33}^{\prime D}$ was chosen to avoid
too large effects in $T_{33}^D$.

Concerning the numerical analysis of a variation of $T_{33}^{\prime D}$, 
the results in the $M_h^{125}$ and the $M_h^{125}(\tilde\tau)$
scenario, are effectively identical due to their identical settings in
the scalar quark sector. The results in the $M_h^{125}(\tilde\chi)$
scenario, however, can vary visibly from the other two scenarios.
For $\de\Mh$ the NH contributions
yield corrections that are in general found to be very small,
and at a similar level of the $T_{33}^{\prime U}$ variation~\cite{Rehman:2022ydc}
(but contrary to previous claims in the literature).
They reach up to $\sim \pm 60 \mev$ in the analyzed parameter space, where
P4 exhibits the largest corrections.
Since the corrections turn out to be very small over the whole
analyzed parameter space we find that the NH terms {\em do not}
alleviate the fact that large SUSY masses are needed to reach the value
of $\Mh \sim 125 \gev$. 
The numerical effects for $\MH$ and $\MHp$ were found to be mostly 
negative and reached values of up to $\sim -60 \gev$ for $\MH$ and $\MHp$. 

Concerning the analysis of a variation of $\mu^\prime$, the
size of the numerical effects depend on the choice of the adjustment of
the holomorphic SSB terms. They can go down to $\de\Mh \sim -1 \gev$ in
the $M_h^{125}$ and $M_h^{125}(\tilde\chi)$ scenarios, and due to a
specific definition of the scalar lepton sector down to $\sim -15 \gev$
in the $M_h^{125}(\tilde\tau)$ scenario. However, those parameter points would
then yield a light $\cp$-even Higgs-boson mass well below $125 \gev$,
i.e.\ they would be experimentally excluded. The numerical effects on
$\MH$ range roughly between $-40 \gev$ and $+40 \gev$, while the effects
on $\MHp$ are mostly negative, going down to $\sim -20 \gev$ in the 
$M_h^{125}$ and the $M_h^{125}(\tilde\tau)$ scenario, and down to
$\sim -40 \gev$ in the $M_h^{125}(\tilde\chi)$ scenario. 

Despite the fact that the NH contributions entering via $T_{33}^{\prime D}$
and $\mu^\prime$ are mostly small for $\Mh$ (the same holds for
$T_{33}^{\prime U}$), a full analysis of
supersymmetric extensions of the SM should 
include the possibility of NH contributions. Consequently,
we aim for an inclusion of these effects into the code
\fh\cite{mhiggslong, mhiggsAEC, mhcMSSMlong, Frank:2006yh, Mh-logresum,
  Bahl:2016brp, Bahl:2017aev, Bahl:2018qog}, thus providing the NHSSM
mass spectrum with the highest possible precision.

\subsection*{Acknowledgments}
S.H.\ acknowledges partial financial support by the Spanish
Research Agency (Agencia Estatal de Investigaci\'on) through the grant
IFT Centro de Excelencia Severo Ochoa No CEX2020-001007-S funded by
MCIN/AEI/10.13039/501100011033. 
The work of S.H.\ was also supported by the Grant PID2022-142545NB-C21
funded by MCIN/AEI/10.13039/501100011033/ FEDER, UE.


\newpage
\pagebreak


\clearpage
\begin{thebibliography}{99}
\bibitem{Fayet:1974pd}
P.~Fayet,
Nucl. Phys. B \textbf{90} (1975), 104-124

\bibitem{Fayet:1976et}
P.~Fayet,
Phys. Lett. B \textbf{64} (1976), 159

\bibitem{Fayet:1977yc}
P.~Fayet,
Phys. Lett. B \textbf{69} (1977), 489

\bibitem{Nilles:1983ge}
H.~P.~Nilles,
Phys. Rept. \textbf{110} (1984), 1-162

\bibitem{Haber:1984rc}
H.~E.~Haber and G.~L.~Kane,
Phys. Rept. \textbf{117} (1985), 75-263

\bibitem{Barbieri:1987xf}
R.~Barbieri,
Riv. Nuovo Cim. \textbf{11N4} (1988), 1-45

\bibitem{Sekmen:2022vzu}
S.~Sekmen [ATLAS, CMS and LHCb],
[arXiv:2204.03053 [hep-ex]].

\bibitem{Girardello:1981wz}
L.~Girardello and M.~T.~Grisaru,
Nucl. Phys. B \textbf{194} (1982), 65
%

\bibitem{Bagger:1993ji}
J.~Bagger and E.~Poppitz,
Phys. Rev. Lett. \textbf{71} (1993), 2380-2382
[arXiv:hep-ph/9307317 [hep-ph]].
%

\bibitem{Chattopadhyay:2017qvh}
U.~Chattopadhyay, D.~Das and S.~Mukherjee,
JHEP \textbf{01} (2018), 158
[arXiv:1710.10120 [hep-ph]].

\bibitem{Un:2023wws}
C.~S.~Un,
Turk. J. Phys. \textbf{48} (2024) no.1, 1-27
[arXiv:2308.12862 [hep-ph]].

\bibitem{Chattopadhyay:2018tqv}
U.~Chattopadhyay, A.~Datta, S.~Mukherjee and A.~K.~Swain,
JHEP \textbf{10} (2018), 202
[arXiv:1809.05438 [hep-ph]].

\bibitem{Chattopadhyay:2019ycs}
U.~Chattopadhyay, D.~Das and S.~Mukherjee,
JHEP \textbf{06} (2020), 015
[arXiv:1911.05543 [hep-ph]].

\bibitem{Chattopadhyay:2022ecq}
U.~Chattopadhyay, A.~Datta, S.~Mukherjee and A.~K.~Swain,
JHEP \textbf{08} (2022), 113
[arXiv:2201.00621 [hep-ph]].

\bibitem{Chakrabortty:2011zz}
J.~Chakrabortty and S.~Roy,
Phys. Rev. D \textbf{85} (2012), 035014
[arXiv:1104.1387 [hep-ph]].

\bibitem{Israr:2024ubp}
S.~Israr and M.~Rehman,
[arXiv:2407.01210 [hep-ph]].

\bibitem{Rehman:2024tdr}
M.~Rehman and S.~Heinemeyer,
[arXiv:2411.00479 [hep-ph]].

\bibitem{Israr:2025cfd}
S.~Israr, M.~E.~G\'omez and M.~Rehman,
Particles \textbf{8} (2025) no.1, 30

\bibitem{Jack:1999ud}
I.~Jack and D.~Jones,
Phys. Lett. B \textbf{457} (1999), 101-108
[arXiv:hep-ph/9903365 [hep-ph]].

\bibitem{Jack:1999fa}
I.~Jack and D.~Jones,
Phys. Rev. D \textbf{61} (2000), 095002
[arXiv:hep-ph/9909570 [hep-ph]].

\bibitem{Jack:2004dv}
I.~Jack, D.~Jones and A.~Kord,
Phys. Lett. B \textbf{588} (2004), 127-135
[arXiv:hep-ph/0402045 [hep-ph]].

\bibitem{Cakir:2005hd}
M.~Cakir, S.~Mutlu and L.~Solmaz,
Phys. Rev. D \textbf{71} (2005), 115005
[arXiv:hep-ph/0501286 [hep-ph]].

\bibitem{Sabanci:2008qp}
A.~Sabanci, A.~Hayreter and L.~Solmaz,
Phys. Lett. B \textbf{661} (2008), 154-157
[arXiv:0801.2029 [hep-ph]].

\bibitem{Un:2014afa}
C.~S.~Un, S.~H.~Tanyıldızı, S.~Kerman and L.~Solmaz,
Phys. Rev. D \textbf{91} (2015) no.10, 105033
[arXiv:1412.1440 [hep-ph]].

\bibitem{Chattopadhyay:2016ivr} 
  U.~Chattopadhyay and A.~Dey,
  JHEP {\bf 1610}, 027 (2016)

\bibitem{Rehman:2022ydc}
M.~Rehman and S.~Heinemeyer,
Phys. Rev. D \textbf{107} (2023) no.9, 095033
[arXiv:2212.13757 [hep-ph]].

\bibitem{Porod:2003um}
W.~Porod,
Comput. Phys. Commun. \textbf{153} (2003), 275-315
[arXiv:hep-ph/0301101 [hep-ph]].

\bibitem{Staub:2009bi}
F.~Staub,
Comput. Phys. Commun. \textbf{181} (2010), 1077-1086
[arXiv:0909.2863 [hep-ph]].

\bibitem{Staub:2010jh}
F.~Staub,
Comput. Phys. Commun. \textbf{182} (2011), 808-833
[arXiv:1002.0840 [hep-ph]].

\bibitem{Staub:2012pb}
F.~Staub,
Comput. Phys. Commun. \textbf{184} (2013), 1792-1809
[arXiv:1207.0906 [hep-ph]].

\bibitem{Staub:2013tta} 
  F.~Staub,
  Comput.\ Phys.\ Commun.\  {\bf 185}, 1773 (2014)
  [arXiv:1309.7223 [hep-ph]].

\bibitem{Staub:2015kfa}
F.~Staub,
Adv. High Energy Phys. \textbf{2015} (2015), 840780
[arXiv:1503.04200 [hep-ph]].

\bibitem{Draper:2016pys}
P.~Draper and H.~Rzehak,
Phys. Rept. \textbf{619}, 1-24 (2016)
[arXiv:1601.01890 [hep-ph]].

\bibitem{Slavich:2020zjv}
P.~Slavich and S.~Heinemeyer (eds), E.~Bagnaschi, H.~Bahl, M.~Goodsell, H.~E.~Haber, T.~Hahn, R.~Harlander, W.~Hollik and G.~Lee, \textit{et al.}
Eur. Phys. J. C \textbf{81} (2021) no.5, 450
[arXiv:2012.15629 [hep-ph]].

\bibitem{Heinemeyer:2011aa}
S.~Heinemeyer, O.~Stal and G.~Weiglein,
Phys. Lett. B \textbf{710} (2012), 201-206
[arXiv:1112.3026 [hep-ph]].

\bibitem{ParticleDataGroup:2024cfk}
S.~Navas \textit{et al.} [Particle Data Group],
Phys. Rev. D \textbf{110} (2024) no.3, 030001

\bibitem{Cepeda:2019klc}
M.~Cepeda, S.~Gori, P.~Ilten, M.~Kado, F.~Riva, R.~Abdul Khalek, A.~Aboubrahim, J.~Alimena, S.~Alioli and A.~Alves, \textit{et al.}
CERN Yellow Rep. Monogr. \textbf{7} (2019), 221-584
[arXiv:1902.00134 [hep-ph]].

\bibitem{Bambade:2019fyw}
P.~Bambade, T.~Barklow, T.~Behnke, M.~Berggren, J.~Brau, P.~Burrows, D.~Denisov, A.~Faus-Golfe, B.~Foster and K.~Fujii, \textit{et al.}
[arXiv:1903.01629 [hep-ex]].

\bibitem{deBlas:2019rxi}
J.~de Blas, M.~Cepeda, J.~D'Hondt, R.~K.~Ellis, C.~Grojean, B.~Heinemann, F.~Maltoni, A.~Nisati, E.~Petit and R.~Rattazzi, \textit{et al.}
JHEP \textbf{01} (2020), 139
[arXiv:1905.03764 [hep-ph]].

\bibitem{Gennai:2007ys}
S.~Gennai, S.~Heinemeyer, A.~Kalinowski, R.~Kinnunen, S.~Lehti, A.~Nikitenko and G.~Weiglein,
Eur. Phys. J. C \textbf{52} (2007), 383-395
[arXiv:0704.0619 [hep-ph]].

\bibitem{Frank:2013hba}
M.~Frank, L.~Galeta, T.~Hahn, S.~Heinemeyer, W.~Hollik, H.~Rzehak and G.~Weiglein,
Phys. Rev. D \textbf{88} (2013) no.5, 055013
[arXiv:1306.1156 [hep-ph]].

\bibitem{Beuria:2017gtf}
J.~Beuria and A.~Dey,
JHEP \textbf{10} (2017), 154
[arXiv:1708.08361 [hep-ph]].

\bibitem{Staub:2011dp}
F.~Staub, T.~Ohl, W.~Porod and C.~Speckner,
Comput. Phys. Commun. \textbf{183} (2012), 2165-2206
[arXiv:1109.5147 [hep-ph]].

\bibitem{Bahl:2018zmf}
E.~Bagnaschi, H.~Bahl, E.~Fuchs, T.~Hahn, S.~Heinemeyer, S.~Liebler, S.~Patel, P.~Slavich, T.~Stefaniak, C.~E.~Wagner and G.~Weiglein,
Eur. Phys. J. C \textbf{79} (2019) no.7, 617
[arXiv:1808.07542 [hep-ph]].

\bibitem{ATLAS:2020zms}
G.~Aad \textit{et al.} [ATLAS],
Phys. Rev. Lett. \textbf{125} (2020) no.5, 051801
[arXiv:2002.12223 [hep-ex]].

\bibitem{CMS:2022goy}
 [CMS],
[arXiv:2208.02717 [hep-ex]].

\bibitem{mhiggslong} S.~Heinemeyer, W.~Hollik and G.~Weiglein,
                    {\em Eur. Phys. J.} {\bf C 9} (1999) 343
                    [arXiv:hep-ph/9812472].

\bibitem{mhiggsAEC} G.~Degrassi, S.~Heinemeyer, W.~Hollik,
                   P.~Slavich and G.~Weiglein,
                   {\em Eur. Phys. J.} {\bf C 28} (2003) 133
                   [arXiv:hep-ph/0212020].

\bibitem{mhcMSSMlong}
                   M.~Frank, T.~Hahn, S.~Heinemeyer, W.~Hollik, 
                   R.~Rzehak and G.~Weiglein,
                   {\em JHEP} {\bf 0702} (2007) 047
                   [arXiv:hep-ph/0611326].

\bibitem{Frank:2006yh} 
  M.~Frank, T.~Hahn, S.~Heinemeyer, W.~Hollik, H.~Rzehak and G.~Weiglein,
  JHEP {\bf 0702}, 047 (2007)
  [hep-ph/0611326].

\bibitem{Mh-logresum}  T.~Hahn, S.~Heinemeyer, W.~Hollik, H.~Rzehak and
  G.~Weiglein,
  {\em Phys.\ Rev.\ Lett.} {\bf 112} (2014) 141801
  [arXiv:1312.4937 [hep-ph]].

\bibitem{Bahl:2016brp}
H.~Bahl and W.~Hollik,
Eur. Phys. J. C \textbf{76} (2016) no.9, 499
[arXiv:1608.01880 [hep-ph]].

\bibitem{Bahl:2017aev}
H.~Bahl, S.~Heinemeyer, W.~Hollik and G.~Weiglein,
Eur. Phys. J. C \textbf{78} (2018) no.1, 57
[arXiv:1706.00346 [hep-ph]].

\bibitem{Bahl:2018qog}
H.~Bahl, T.~Hahn, S.~Heinemeyer, W.~Hollik, S.~Paßehr, H.~Rzehak and G.~Weiglein,
Comput. Phys. Commun. \textbf{249} (2020), 107099
[arXiv:1811.09073 [hep-ph]].
\end{thebibliography}
\end{document}